\documentclass[pre,twocolumn,showpacs,preprintnumbers,amsmath,amssymb]{revtex4} 
\usepackage{graphicx}
\begin{document}
\title{Analytical investigation of oscillations in intersecting flows of pedestrian and vehicle traffic} 
\author{Dirk Helbing, Rui Jiang, and Martin Treiber} 
\affiliation{Dresden University of Technology, Andreas-Schubert-Str. 23, 01062 Dresden, Germany}
\begin{abstract} 
In two intersecting many-particle streams, one can often find the emergence of oscillatory
patterns. Here, we investigate the interaction of pedestrians with vehicles, when they try to
cross a road. A numerical study of this coupled pedestrian-vehicle delay problem has been 
presented in a previous paper. Here, we focus on the analytical treatment of the problem,
which requires to use a simplified car-following model. Our analytical results for the phase transition
to oscillatory pedestrian and traffic flows and the average waiting times are well 
supported by numerical evaluations and give a detailed picture of the collective dynamics
emerging when pedestrians try to cross a road. The mathematical expressions allow one to
identify the dependence on model parameters such as the vehicle or pedestrian arrival rate, and
the safety factor of pedestrian gap acceptance. We also calculate a formula for the vehicle
time gap distribution, which corresponds to the departure time distribution of a M/D/1 queue.
\end{abstract}
\maketitle

\section{Introduction}

Pattern formation is a wide-spread feature of driven many-particle systems.
In particular, oscillatory patterns are found in fluids, granular materials, colloidal
systems, and traffic flows. One typical example are stop-and-go waves in
traffic flows on freeways caused by a delayed adaptation to changing traffic conditions
\cite{rev1,rev2,rev3,rev4,rev5}. Emergent oscillations have been discovered in so different systems as the
density oscillator \cite{Steinbock}, ticking hour glass \cite{ticking}, RNA Polymerase traffic on DNA
\cite{Sneppen}, pedestrians passing a bottleneck \cite{ped,panic}, or 
ants \cite{Dussu}. Oscillatory patterns have also been found in two intersecting
pedestrians streams \cite{transci} or simulations of colloidal systems \cite{Loewen}.
\par
Although the subject is rather old \cite{1,2,3}, the crossing of vehicle streams by pedestrians 
has recently attracted an increasing interest, also among physicists \cite{4,5,6,7,numerics}.
However, the problem of interactions between vehicles and pedestrians, when pedestrians are trying
to cross a road, has not yet been sufficiently understood. The mathematical investigation of this
problem will be the subject of this paper. Numerical studies have shown
a transition from crossing the road one by one or in small groups to 
coupled oscillations of pedestrian and vehicle flows, if pedestrians use small gaps
to cross the road \cite{numerics}. In the following, the dynamics of this phenomenon and the 
parameter-dependence of the transition point will be investigated analytically.
\par
Our paper is organized as follows: Section~\ref{model} formulates the model for the pedestrian and
vehicle behavior and their interactions. Moreover, we calculate a formula for an idealized
vehicle time gap distribution. In Sec.~\ref{analyt}, we will derive analytical results
on the dynamic behavior of interacting pedestrian and vehicle flows. Moreover, we will compare these
results with numerical evaluations of computer simulations of the underlying model. 
Our analytical formulas for the transition point and the waiting times of pedestrians and cars
are well compatible with numerically determined data. Finally, we will
summarize and discuss our results in Sec.~\ref{summa}, which are relevant for many
systems with intersecting flows or competing processes.

\section{Formulation of the Model} \label{model}
\subsection{Vehicle Behavior}\label{veh}

In our simplified model of vehicle dynamics, cars are treated as moving objects of length $l_0$.
We assume a constant arrival flow $Q_{\rm arr}$ of vehicles and that new cars try to enter the 
investigated road section with a probability $q=Q_{\rm arr} \, dt$ per time step $dt$. 
This implies an exponential time gap distribution, which is modified by vehicle-vehicle 
interactions (see Sec.~\ref{Dis}).
In fact, a vehicle with the speed $v$ following a leading vehicle with speed $v_*$ is assumed to
decelerate with $dv/dt = -a$, if $v>0$ and
\begin{equation}
 \Delta x < l_0 + d_0 + \frac{v^2}{2a} - \frac{v_*{}^2}{2a} \, ,
\label{greater}
\end{equation}
where $\Delta x$ denotes the distance between the two vehicles, $l_0$ the vehicle length and $d_0$ 
the preferred minimum bumper-to-bumper distance among
cars. This condition guarantees accident-free driving \cite{Krauss}.
For a ``$>$''-sign in Eq.~(\ref{greater}), the vehicle accelerates with
$dv/dt =a$, delayed by the reaction time $T$, 
until the maximum (free) speed $v_0$ is reached. For an ``$=$''-sign in Eq.~(\ref{greater}), the
velocity is not changed, i.e. $dv/dt =0$. The above continuous car-following model may be called
the constant-deceleration-delayed-acceleration model (CDDA model) and has some
similarities with the slow-to-start cellular automaton model \cite{slow}.
\par
 We assume that pedestrians enter the street at the crossing point
O, when they consider it safe (see Sec.~\ref{ped}). Moreover, crossing the road takes 
a time period $\tau$.
In order to avoid accidents with pedestrians, vehicles decelerate with $dv/dt = -a$ if necessary. We
consider two different deceleration rules:
\begin{itemize}
\item[a)] {\bf Careful drivers:} The closest car to a pedestrian on the street decelerates, 
if the distance $d(t) = -x(t)$ to the crossing point O is within the range
\begin{equation}
 0 \le d(t) \le d_0 + \frac{v{}^2}{2a} \,
\end{equation}
where $d_0$ is the safety distance that a car should keep from a crossing pedestrian. 
We assumed this safety distance to be identical to the minimum
bumper-to-bumper distance among vehicles appearing in (\ref{greater}).
\item[b)] {\bf Aggressive drivers:} The closest car starts to decelerate at the time $t_0$ determined
so that the distance to the pedestrian corresponds to the safety distance
$d(t_n+ \tau) = d_0$ at the time $t_n+\tau$ when the last (the $n$-th) pedestrian on the street 
(entering at time $t_n$) leaves the road after the crossing time $\tau$.
\end{itemize}
After the last pedestrian has left the street, i.e. at time $t_n+\tau$, the car accelerates with $dv/dt = a$, until
it has reached its desired velocity $v_0$ again. The characteristic distance between stopped vehicles 
in a queue is the vehicle length $l_0$ plus the minimum bumper-to-bumper distance $d_0$, 
which defines the jam density
\begin{equation}
 \rho_{\rm jam} := \frac{1}{l_0+d_0} \, .
\end{equation}
In the following, we will assume that a car starts to accelerate after its leader delayed by 
the reaction time $T$. This implies that the following car has reached the position
of the leading car in the queue after a time period 
$T + \sqrt{2(d_0+l_0)/a }$ and that the 
distance to the leading car is $l_0 + d_0 + Tv_0$, when the following car has reached its 
maximum velocity $v_0$. Therefore, the outflow from a traffic jam starts with a value of
$( T + \sqrt{2(d_0+l_0)/a })^{-1}$ and eventually reaches the characteristic (maximum) value 
\begin{equation}
 Q_{\rm out} := \left( T + \frac{l_0+d_0}{v_0} \right)^{-1} \, , 
\end{equation}
while the traffic jam (queue) resolves upstream with the characteristic speed 
\begin{equation}
 c := \frac{l_0+d_0}{T} = \frac{1}{\rho_{\rm jam} T}  
\end{equation}
due to the distance $l_0+d_0$ between queued cars and the delay $T$ in acceleration.
Moreover, when a vehicle is stopped at point $x(t) = - d_0$, the forming traffic jam behind it propagates
upstream with the velocity \cite{control}
\begin{equation}
 C := \left( \frac{\rho_{\rm jam}}{Q_{\rm arr}} - \frac{1}{v_0} \right)^{-1} \, ,
\label{propagates}
\end{equation}
which depends on the vehicle arrival rate $Q_{\rm arr}$.
\par
The proposed simple 
car-following model essentially reflects the features of the section-based, fluid-dynamic
traffic flow model proposed in Ref.~\cite{control}, with the only
difference that the acceleration and braking processes 
require time periods of $T+v_0/a$ and $T/v_0$, respectively. 
Apart from scaling time and space variables in order to get rid of two more
model parameters, it is hard to think of any further simplification of the above vehicle model without sacrificing
fundamental properties of traffic flows such as the constant outflow from
traffic jams and the characteristic jam resolution speed \cite{Kerner}.
Nevertheless, it may be interesting to study 
the limit $a \rightarrow \infty$ of unlimited acceleration possibilities, which eliminates
acceleration and deceleration times. More realistic variants of the above car-following 
model, however, should distinguish different acceleration and deceleration strengths $a$ and $b$, which have
been set equal here for the sake of simplicity. A stochastic variant of this 
model describing a fluctuating acceleration behavior 
would be also interesting to study.

\subsection{Idealized Vehicle Distance Distribution} \label{Dis}

In our vehicle simulations, we have generated vehicles with initial velocity $v=0$ at the upstream boundary of the simulation
stretch according to the exponential time gap distribution $Q_{\rm arr} \mbox{e}^{-Q_{\rm arr}\,T'}$,
where $T'$ denotes the actual time gap. However, according to our car-following model, vehicles had gained at least 
their preferred distance $D= l_0+d_0+v_0T$, when they reached the maximum speed $v_0$. According
to theoretical considerations, this changed the effective time-gap distribution at the crossing point to 
\begin{eqnarray}
 P(T') &=& Q_{\rm arr}T_0 \, \delta(T'-T_0) \nonumber \\
&+& (1-Q_{\rm arr} T_0)  Q_{\rm arr} \mbox{e}^{-Q_{\rm arr}(T'-T_0)} 
 \Theta(T' - T_0) \qquad
\label{resu}
\end{eqnarray}
with $T_0=D/v_0$ (see Appendix), when no vehicles at the entry point were dropped. 
That is, a fraction $Q_{\rm arr}T_0$ of vehicles will follow with the desired time gap $T_0$,
while the rest has an exponentially distributed, larger time gap $T'>T_0$. 
$\delta(y)$ denotes Dirac's delta function, while the Heaviside function $\Theta(y)$ is 1 for
$y\ge 0$ and 0 otherwise.
\par
Our exponentially distributed
vehicle generation mechanism sometimes causes a virtual queue of vehicles at the upstream boundary, which 
can be avoided by generating vehicles according to the resulting time gap distribution (\ref{resu}).
In fact, our implementation of the boundary conditions corresponds to a $M/D/1$ queuing system \cite{MD1,MD2},
i.e. to a queue with Poissonian distributed Markovian arrivals (where the time gaps between successive arrivals
are exponentially distributed), while the service rate $1/T_0$ is assumed to be deterministic. (The ``1'' stands for one 
``channel'', i.e. no parallel service.) 
\par
Now, let $P_0$ be the probability that no vehicle is waiting in the queue
to be served, i.e. to enter the road. The probability of releasing the next vehicle with a time gap $T'=T_0$ is then given
by the probability $(1-P_0)$ of having queued vehicles waiting to enter, plus the probability 
$P_0(1-\mbox{e}^{-Q_{\rm arr}T_0})$ that we have the no-queue case and a vehicle arrives 
during the service time $T_0$. In cases with no queue
where the time gap $T'$ of the next arriving vehicle is greater than $T_0$, we have an exponential
time gap distribution $Q_{\rm arr}\,\mbox{e}^{-Q_{\rm arr}T'}/\mbox{e}^{-Q_{\rm arr}T_0}$,
where $\mbox{e}^{-Q_{\rm arr}T_0}$ is the normalization factor of the conditional probability of finding
time gaps larger than $T_0$. Altogether, we obtain the time gap distribution 
\begin{eqnarray}
 P(T') &=& [(1-P_0) + P_0 (1- \mbox{e}^{-Q_{\rm arr}T_0})] \, \delta(T' - T_0) \nonumber \\
 &+& P_0  Q_{\rm arr}\,\mbox{e}^{-Q_{\rm arr}T'} \Theta(T' - T_0) \, .
\end{eqnarray}
Demanding 
\begin{equation} 
 \frac{1}{Q_{\rm arr}} = \int\limits_0^\infty \, dT' T' P(T') 
 = T_0 + \frac{P_0}{Q_{\rm arr}} \, \mbox{e}^{-Q_{\rm arr}T_0} \, , 
\end{equation} 
i.e. that the vehicle flow $Q_{\rm arr}$ and, therefore, the average time gap remains unchanged,
we find
\begin{equation}
 P_0 = ( 1 - Q_{\rm arr} T_0)\, \mbox{e}^{Q_{\rm arr}T_0} \, .
\end{equation}
This implies the idealized vehicle time gap distribution (\ref{resu}), which will be necessary to
evaluate the expected waiting time of pedestrians for a suitable time gap to cross the road (see Appendix).

\subsection{Pedestrian Behavior}\label{ped}

We will assume that pedestrians enter the sidewalk of the street at the crossing point O with 
probability $p = \lambda \, dt$ per time step $dt$, i.e. $\lambda$ denotes the arrival rate of pedestrians.
If there is no sufficient gap in the vehicle stream to cross, they accumulate around point O, but
they start immediately to enter the road at time $t$, if $v(t) = 0$
(i.e. if the vehicle velocity is zero) or if
\begin{equation}
  d(t) > d_0  \quad \mbox{and} \quad \Delta t(t) := \frac{d(t)}{v(t)} \ge \sigma \tau
\label{criterion}
\end{equation}
(i.e. if the distance $d(t)$ is larger than the preferred safety distance $d_0$
and the time gap $\Delta t$ is large enough to cross the road).
Here, $\Delta t$ is the time to collision of the nearest approaching vehicle and $\sigma$ a safety factor
of pedestrians. $\tau$ is the time period required for a pedestrian to cross (one lane of) the road.
We may distinguish two limiting cases of gap selection, i.e. interactions with approaching vehicles:
\begin{itemize}
\item[1)] {\bf Careful pedestrians} assume that cars may not decelerate and approach with their desired
velocity $v_0$. They cross the road only, if the car at no time comes closer than the preferred safety distance
$d_0$, which implies the following choice of the safety factor:
\begin{equation}
 \sigma = \sigma_1 := 1 + \frac{d_0}{v_0 \tau} \, .
\end{equation} 
\item[2)] {\bf Daring pedestrians} enter the road, if a car with velocity $v_0$ would not come
closer than the preferred safety distance $d_0$, if it decelerated with $dv/dt = -a$ in order to avoid an
accident. This implies the reduced safety factor
\begin{equation}
 \sigma = \sigma_2 := 1  + \frac{d_0}{v_0 \tau} - \frac{a\tau}{2v_0} = \sigma_1 - \frac{a\tau}{2v_0} \, .
\label{sig2}
\end{equation}
In this case, a single pedestrian can force a car to stop, namely when entering at a vehicle distance
$d(t) = d_0 + v_0{}^2/(2a)$.
\end{itemize}
Realistic values of the safety factor $\sigma$ are expected to be above $\sigma_2$. 
\par
For the following analysis,
we will identify the time point $t=0$ with the time when the first pedestrian(s), who cause(s) a vehicle to decelerate,
enter(s) the road. The entering time of the next entering pedestrian is denoted by $t_1$, the entering time of
the $k$th following pedestrian by $t_k$, and the entering time of the last ($n$th) following pedestrian
before the car passes point O by $t_n$. 

\subsection{Simulated Dynamic Behavior of Interacting Vehicle and Pedestrian Flows}

\begin{figure}
\begin{center}
\includegraphics[width=8cm]{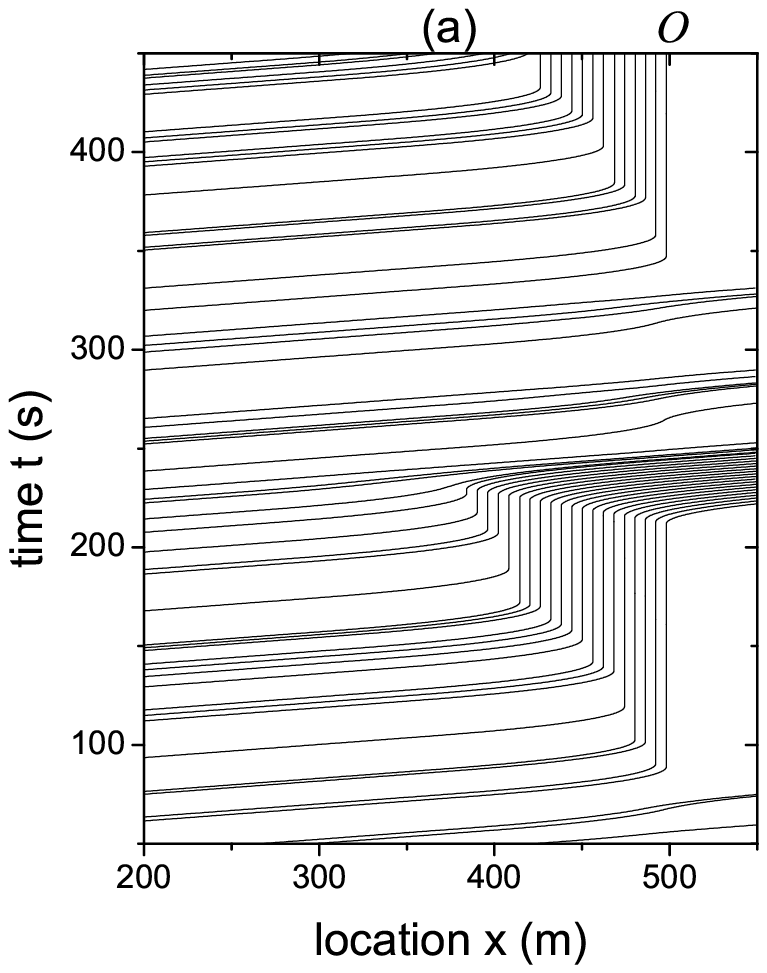}
\includegraphics[width=8cm]{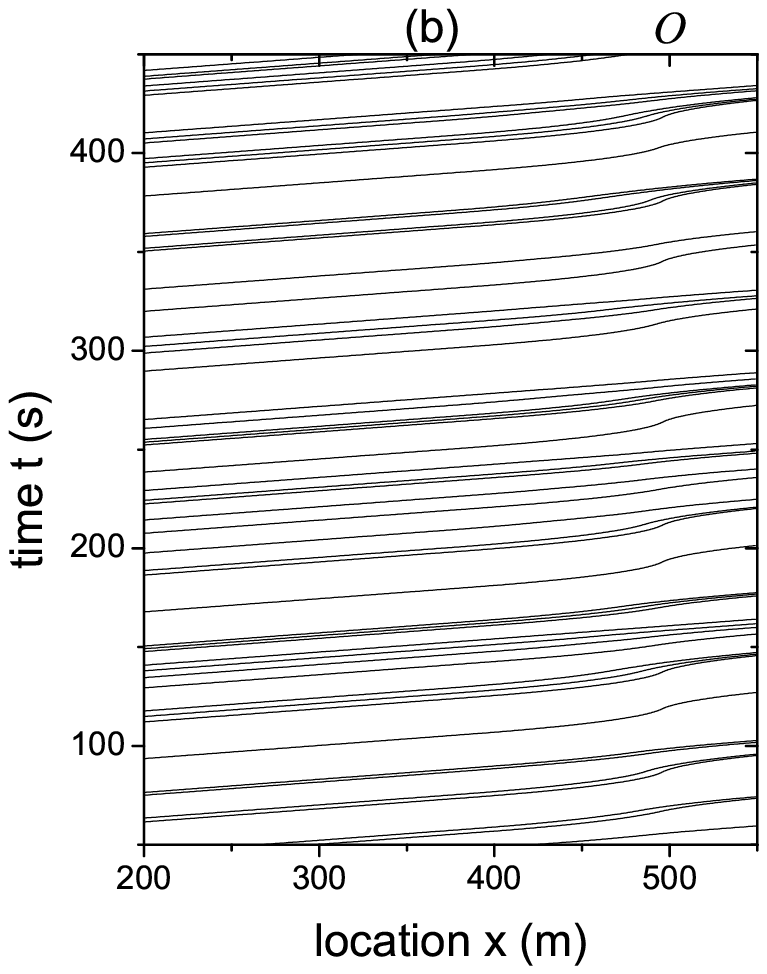} 
\end{center}
\caption[]{(a) Representative space-over-time plot of vehicle trajectories for careful drivers and
the pedestrian safety factor
$\sigma=1.05$. Pedestrians may stop cars, which causes vehicle queues. These suppress 
the crossing of newly arriving pedestrians until the vehicle queue has completely dissolved. 
(b) Representative space-over-time plot of vehicle trajectories for the larger safety factor 
$\sigma=1.25$, for which pedestrians use large gaps only. As a consequence, pedestrians 
do not stop cars completely when they 
cross the street, and no vehicle queues are formed.\label{fig1}}
\end{figure}

Simulations of vehicles interacting with pedestrians crossing a street have recently shown an
interesting phenomenon: While for large enough values of the safety factor $\sigma$, pedestrians
cross the road one by one or in small groups, one finds alternating pedestrian and vehicle streams
if the safety factor is smaller than some critical value $\sigma_0$. This value can be exactly 
calculated for the above model (see Eq.~(\ref{sigma0})), 
which shows qualitatively the same dynamic behavior like the variant of the IDM
model studied in a previous publication \cite{numerics}. Representative simulation results for the above
proposed pedestrian and vehicle model are displayed in Fig.~\ref{fig1}. The parameter values
used in this paper are $a=1$~m/s$^2$, $\tau=2$~s, $T=0.9$~s, $l_0=4$~m, $d_0 = 2$~m, 
and $v_0 =15$~m/s, and our numerical investigation focusses on careful drivers.
\par
The reason for the observed oscillations is that pedestrians can force vehicles to stop, if they choose
small time gaps $\Delta t$. However, if vehicles are stopped, they have to wait until there is a gap of period
$\tau$ or larger in the pedestrian stream, before they can accelerate again. During this waiting time,
a vehicle queue is formed, which can become very long, dependent on the vehicle arrival rate. 
Pedestrians cannot cross the road again, before this queue is completely dissolved, at least if 
\begin{equation}
   \sigma \tau > \frac{d_0 + T v_0}{v_0} = T + \frac{d_0}{v_0} \, ,
\end{equation}
i.e. if the time gap between successive vehicles having left the queue
is too short for pedestrians to enter the street, and if
\begin{equation}
 \sigma \tau > \frac{l_0 + 2d_0  - \frac{a}{2} \left( \sqrt{\frac{2(l_0+d_0)}{a}} - T \right)^2}
 {a \left( \sqrt{\frac{2(l_0+d_0)}{a}} - T \right)} \, ,
\end{equation}
i.e. if the time gap with respect to the second car in the queue at the time $\sqrt{2(l_0+d_0)/a}>T$
(when the back of the first vehicle has passed the crossing point O) is not large enough for
pedestrians to enter the street.
\par 
In summary, we may 
have alternating time periods in which pedestrians can cross the road and time periods in which cars can
pass point O. In the following sections, based on statistical approaches, 
we will try to estimate the time period until a sufficiently large gap in the vehicle flow occurs to allow
pedestrians a crossing of the road. Likewise, we will calculate the time period 
until queued vehicles find a large enough gap between crossing pedestrians, allowing them to accelerate again.
Analytical results can be only gained for simple models as the one proposed above. Nevertheless,
we expect qualitatively similar relationships for a broad class of other traffic models.

\section{Analytical Results and Comparison with Computer Simulations}\label{analyt}

\subsection{Dynamics of Vehicles Reacting to Pedestrians}\label{ttc}

\begin{figure}
\begin{center}
\includegraphics[width=8cm]{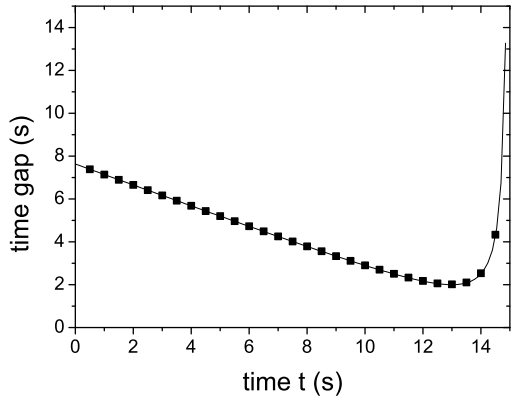}
\end{center}
\caption[]{Time-dependent time to collision $\Delta t(t)=d(t)/v(t)$ for careful drivers [see Eq. (\ref{stop})], 
when pedestrians would enter the road with probability $p=1$ and $\sigma = 1.05$
(symbols = numerically determined values, solid line = analytical formulas).
Due to the braking maneuver, the time to collision goes down in the beginning, but it grows again later on, 
as the vehicle comes to rest at the finite distance $d(t) = d_0$ to the pedestrian.\label{fig2}}
\end{figure}

Let $t_0$ be the time point when the car starts to decelerate as response to a crossing pedestrian.
According to Secs.~\ref{veh} and \ref{ped}, 
we find that the time to collision evolves in time according to
\begin{equation}
 \Delta t(t) = \frac{d(t)}{v(t)} = \frac{d(0)-v_0t}{v_0} = \frac{d(0)}{v_0} - t \quad \mbox{if} \quad t < t_0 \, .
\end{equation}
For careful drivers, i.e. case a), the start time of deceleration can be determined as
\begin{equation}
 t_0 = \frac{d(0)-d_0}{v_0} - \frac{v_0}{2a} \, .
\end{equation}
This yields the time to collision
\begin{eqnarray}
 \Delta t(t) &=& \frac{d_0 + v_0{}^2/(2a) - v_0 (t-t_0) + a(t-t_0)^2/2}{v_0 - a(t-t_0)} \nonumber \\
 &=& \frac{v_0}{2a} - \frac{t-t_0}{2} + \frac{d_0}{v_0 - a(t-t_0)} \mbox{ if } t\ge t_0
\label{stop}
\end{eqnarray}
(see Fig.~\ref{fig2}) and the vehicle velocity
\begin{equation}
 v(t_0+\tau) = v_0 - a(\tau-t_0) = \frac{v_0}{2} - a\tau + a\, \frac{d(0) - d_0}{v_0} 
\end{equation}
after the pedestrian has crossed the road. If the vehicle velocity at the beginning of the braking maneuver
is $v(t_0)< v_0$, one just has to replace $v_0$ by $v(t_0)$. For aggressive drivers, i.e. case b),
we find 
\begin{equation}
 t_0 = \tau - \sqrt{ \frac{2v_0\tau}{a} - 2\, \frac{d(0) - d_0}{a} } \, ,
\end{equation}
\begin{eqnarray}
 \Delta t(t) &=&  \frac{d(t_0) - v_0 (t-t_0) + a(t-t_0)^2/2}{v_0 - a(t-t_0)} \nonumber \\
 &=& \frac{v_0}{2a} - \frac{t-t_0}{2} + \frac{d(t_0) - v_0{}^2/(2a)}{v_0 - a(t-t_0)} \mbox{ if } t\ge t_0,\qquad
\label{go}
\end{eqnarray}
and
\begin{eqnarray}
 v(t_0+\tau) &=& v_0 - a(\tau - t_0) \nonumber \\
&=& v_0 - \sqrt{2av_0 \tau - 2a[d(0)-d_0]} \, .
\end{eqnarray}
That is, the greater the initial distance, the later will the vehicle start to decelerate and the
larger will the resulting velocity be.
Note that, according to the gap acceptance rules of pedestrians outlined in Sec.~\ref{ped}, the
shortest distance to a moving vehicle at which pedestrians enter the road, is given by
$\sigma\tau v$.
\par

\subsection{Average Delay to Vehicles}\label{avdelay}

\begin{figure}
\begin{center}
\includegraphics[width=8cm]{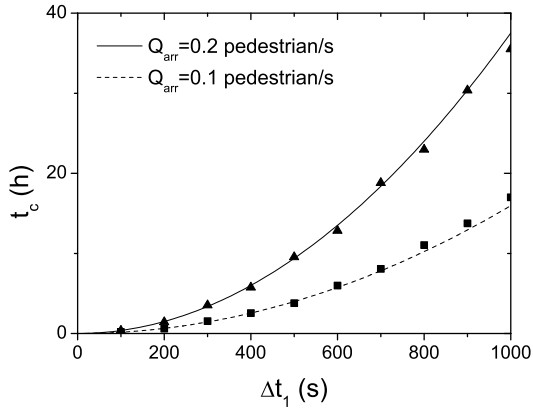}
\end{center}
\caption[]{ Average of the cumulative waiting times $t_{\rm c}$ of vehicles as a function of the time
period $\Delta t_1$ the first vehicle in the queue has to wait, for different values of
the vehicle arrival rate $Q_{\rm arr}$, see Eq. (\ref{propagates}) (symbols = numerically determined values, 
parabolic curves = analytical formula).\label{fig3}}
\end{figure}

Let us denote by $v_{\rm min}$ the minimum velocity before the car accelerates again.
If only one pedestrian obstructs the car, we have $v_{\rm min} = v(\tau)$, as calculated
above. The time delay to the car compared to a movement with the free velocity $v_0$ can
be calculated as the distance $2(v_0-v_{\rm min})^2/(2a)$ travelled less, divided by the
desired velocity $v_0$, which results in
\begin{equation}
 \Delta t_{\rm br} = \frac{(v_0 - v_{\rm min})^2}{a v_0} \, .
\label{vmin}
\end{equation}
If the vehicle is stopped, the time lost by the acceleration and deceleration process amounts to
$v_0/a$. On top of this, we have to add the average waiting time $t_{\rm w}$. 
This can be obtained as follows: If $\Delta t_1$ denotes the waiting time of the first
stopped vehicle, the number of vehicles queuing up behind it until the first car in the queue starts to
accelerate is given by $\rho_{\rm jam} C \, \Delta t_1$. The delay of the last vehicle in
the queue is the queue length $l=C\,\Delta t_1$, divided by the queue resolution
speed $c$. As the waiting time between the first and the last vehicle in the queue progresses approximately
linearly, their cumulative waiting time is given by
\begin{equation}
 \frac{\rho_{\rm jam}C\,\Delta t_1}{2}\left( \Delta t_1 + \frac{C\, \Delta t_1}{c} \right) 
= \frac{\rho_{\rm jam} C (\Delta t_1)^2}{2} \left( 1 + \frac{C}{c} \right) \, .
\label{first}
\end{equation}
Moreover, upto the time point when the queue formed within the stopping time $\Delta t_1$ has resolved,
another $\rho_{\rm jam} l C/(c-C)$ vehicles have joined the queue (cf. Formula (1.48) in 
Ref.~\cite{control}). While the waiting time of the first of these additional vehicles is 
approximately $l/c=C\,\Delta t_1/c$ (as the one of the last vehicle in the first part of the queue),
the waiting time of the last vehicle is basically zero, which implies a cumulative waiting
time of
\begin{equation}
 \frac{\rho_{\rm jam}C\,\Delta t_1}{2}\left( \frac{C}{c-C}\,\frac{C\, \Delta t_1}{c} + 0 \right) 
= \frac{\rho_{\rm jam} C (\Delta t_1)^2}{2} \frac{C^2}{c^2-cC} \, .
\end{equation}
Adding this to (\ref{first}) gives the cumulative waiting time
\begin{equation}
t_{\rm c} = \frac{\rho_{\rm jam}}{2} (\Delta t_1)^2 \frac{c C}{c - C} \, ,
\label{tc}
\end{equation}
which grows quadratically in $\Delta t_1$ (see Fig.~\ref{fig3}).
\par
Finally, dividing this result by the total number $C\,\Delta t_1 [1+C/(c-C)]$
of vehicles yields a very simple relationship for the average waiting time,
which is just given as the average waiting time of the first and the last queued vehicle:
\begin{equation}
 t_{\rm w} = \frac{\Delta t_1}{2} \, .
\end{equation}
However, the estimation of the waiting time $\Delta t_1$ of the first
stopped vehicle is rather difficult (see Sec.~\ref{est}). 

\subsection{Determination of the Transition Point to Alternating Flows}

\begin{figure}
\begin{center}
\includegraphics[width=8cm]{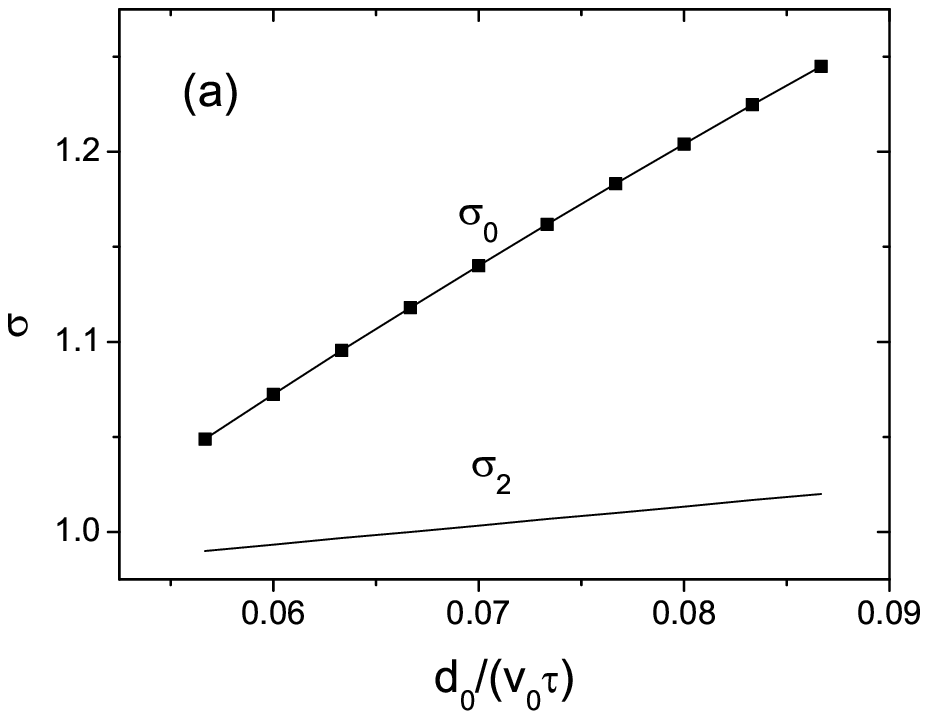}
\includegraphics[width=8cm]{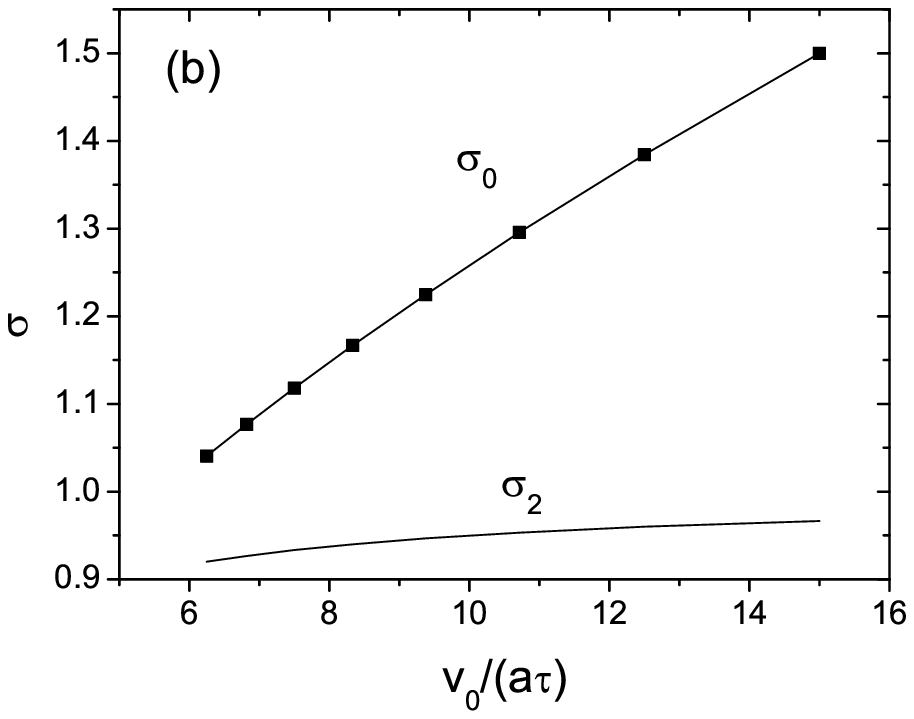}
\includegraphics[width=8cm]{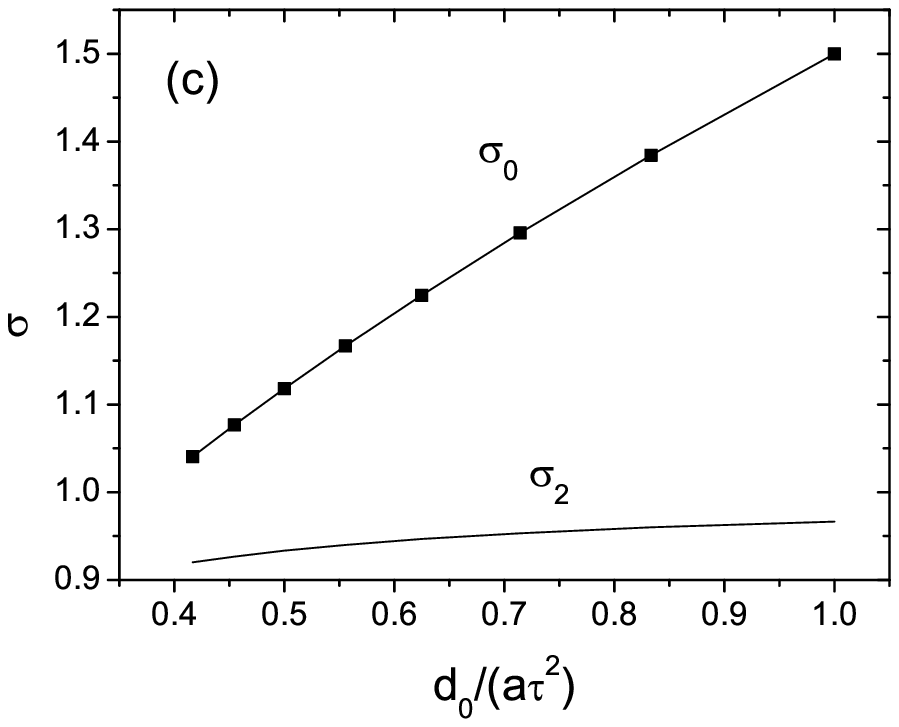}
\end{center}
\caption[]{Transition point $\sigma_0$ to alternating vehicle and pedestrian flows as a function
of the dimensionless parameters (a) $x_1=d_0/(v_0\tau)$ obtained for $d_0 \in [1.6\mbox{ m}, 2.5\mbox{ m}]$,
(b) $x_2=v_0/(a\tau)$ obtained for $a \in [0.5\mbox{ m/s}^2, 1.2\mbox{ m/s}^2]$, and
(c) $x_3 = d_0/(a\tau^2)$ obtained for $a \in [0.5\mbox{ m/s}^2, 1.2\mbox{ m/s}^2]$
in comparison with the lower limit $\sigma_2$ of reasonable safety factors 
[see Eq.~(\ref{sig2})] (symbols = numerically determined values, solid lines 
= analytical formula). Note that the value of $\sigma_0$ is constant for 
$x_3 = x_1x_2 = d_0/(a\tau^2) = \mbox{const.}$\label{fig4}}
\end{figure}

The long vehicle and pedestrian queues required for pronounced
oscillations in the pedestrian and vehicle flows can only occur, if vehicles can
be completely stopped by pedestrians. This cannot happen,  
if the safety factor $\sigma$ of pedestrians is large enough. For small values of $\sigma$, 
however, there exists a time point $t_-$, 
after which the safety criterion (\ref{criterion}) prohibits a further entering of pedestrians into the
road. This time point is given by the earlier time
fulfilling the critical safety condition 
$\Delta t(t_\mp) = \sigma \tau$. Together with the expressions for the times to collision
in Sec.~\ref{ttc}, this eventually implies
\begin{equation}
 t_{\mp} - t_0 = \frac{v_0}{a} - \sigma \tau \mp \sqrt{(\sigma \tau)^2 - \frac{2d_0}{a} } 
\label{tw}
\end{equation}
for careful drivers. $t_+$ is the first time point at which 
pedestrians may re-enter the road again, as 
the time to collision $\Delta t(t)$ increases close to the crossing point [see formula (\ref{stop})].
The car reaches its minimum possible velocity a time period $\tau$ after $t_-$, 
i.e. after the latest entering pedestrian has left the road
at time $t_- + \tau$. With (\ref{tw}) this implies
\begin{equation}
  v(t_-+\tau) = a \tau (\sigma -1) + \sqrt{ (a\sigma \tau)^2 - 2ad_0 } 
\label{minv}
\end{equation}
for careful drivers. For aggressive drivers, we have to replace $d_0$ by $d(t_0) - v_0{}^2/(2a)$.
To exclude stopped vehicles, on the one hand, this minimum velocity should be positive, i.e. 
\begin{equation}
 \left( \sigma - \frac{1}{2}\right) a\tau^2 > d_0 \, .
\label{tog}
\end{equation}
On the other hand, vehicles could also be stopped by new pedestrians entering the road at a
time $t \ge t_+$ that lies before the time $t_-+\tau$ at which the last pedestrian
has left the road. Therefore, in order to avoid the stopping of vehicles by multiple crossing pedestrians,
we have to demand
\begin{equation}
t_+ - t_- = 2\sqrt{(\sigma \tau)^2 - \frac{2d_0}{a} } > \tau \, ,
\end{equation}
which results in 
\begin{equation}
 \sigma > \sqrt{ \frac{2d_0}{a\tau^2} + \frac{1}{4} } \, .
\end{equation}
Together with condition (\ref{tog}) we find that 
a careful driver cannot be stopped completely under the condition
\begin{equation}
\sigma > \sigma_0 = \max \left(\frac{d_0}{a\tau^2} + \frac{1}{2}, \sqrt{\frac{2d_0}{a\tau^2} + \frac{1}{4}}
\right)
\label{sigma0}
\end{equation}
At the value $\sigma = \sigma_0$, we expect a transition from continuous pedestrian
and vehicle flows to alternating flows (see Fig.~\ref{fig4}). 
\par

\subsection{Calculation of Earlier Acceleration}\label{earlier}

\begin{figure}
\begin{center}
\includegraphics[width=8cm]{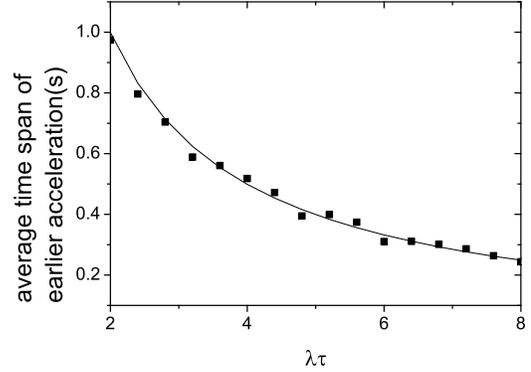}
\end{center}
\caption[]{Average time span $t_- - \langle t_n \rangle$ between the latest {\em possible} entering
of the street by a pedestrian and the time point when the last pedestrian {\em actually} enters the street
as a function of the scaled pedestrian arrival rate $\lambda\tau$, 
see Eq.~(\ref{las}) (symbols = numerically determined values, 
solid line = analytical formula).\label{fig5}}
\end{figure}

Due to the statistical arrival of pedestrians with a rate $\lambda = p/dt$, it is likely that the
time point $t_n\le t_-$ of the last ($n$th) pedestrian entering the road is smaller than the latest {\em possible}
entering time $t_-$. We are, therefore, interested in calculating the mean value $\langle t_- -t_n\rangle
= t_- - \langle t_n \rangle$ of the time gap $t_--t_n$, where $n$ is an arbitrary integer number.
For this, let $K = t_-/dt$ be the number of time steps between the first entering pedestrian
and $t_-$. As the probability that no pedestrian enters in a time step is given by $r=(1-p)$, 
$(1-p)^K$ is the probability that nobody enters between $t=0$ and $t=t_-$, and $p(1-p)^{K-k}$ the probability
that the last pedestrian enters at time $t_- - (K-k)\,dt = k\,dt$. The expected value of $t_- - t_n$ is
\begin{eqnarray}
 \frac{\langle t_--t_n\rangle}{dt} &=& K  (1-p)^N + p \sum_{k=1}^K (K-k)  (1-p)^{K-k} \nonumber \\
 &=& K r^K + (1-r)r \frac{d}{dr} \sum_{k=1}^K r^{K-k} = \frac{r(1-r^K)}{1-r} \nonumber \\
 &=& (1-p) \frac{1- (1-p)^K}{p} 
\end{eqnarray}
(see Fig.~\ref{fig5}).
Therefore, if a vehicle is not stopped, instead of at time $t_-+\tau$,
on average it already starts to accelerate at the earlier time
\begin{eqnarray}
 \langle t_n \rangle +\tau &=& t_-+ \tau -  (1-p)\, dt \frac{1- (1-p)^K}{p} \nonumber \\
&\approx & t_-+ \tau -  \frac{1- \mbox{e}^{-\lambda t_-}}{\lambda} \, ,
\end{eqnarray}
where the last step of this calculation is based on Eq.~(\ref{see}) below. With this result, we can
now estimate the expected value $\langle v_{\rm min} \rangle$
of the minimum vehicle velocity $v_{\rm min}$ entering Eq.~(\ref{vmin}):  
\begin{equation}
 \langle v_{\rm min} \rangle - v(t_-+\tau) = a (t_- - \langle t_n \rangle ) 
 = a (1-\mbox{e}^{-\lambda t_-})/\lambda  \, .
\label{las}
\end{equation}
The higher velocity compared to $v(t_- +\tau)$ given by Eq.~(\ref{minv}) originates from the
earlier car acceleration, i.e. the shorter deceleration time.
\par

\subsection{Estimation of the Waiting Time of the First Vehicle}\label{est}

If a vehicle is stopped by crossing pedestrians after a deceleration time $v_0/a$, it will have to wait
until a time gap of duration $\tau$ in the pedestrian flow occurs. A gap of length
$\tau=N\,dt$ or greater occurs with probability 
\begin{equation}
 (1-p)^N = (1-p)^{\tau/dt} = [ \underbrace{(1-p)^{1/dt}}_{=\,\mbox{e}^{-\lambda}} ]^\tau
 = \mbox{e}^{-\lambda \tau} \, , 
\label{see} 
\end{equation}
i.e. gap sizes are exponentially distributed, as expected.
Here, we have assumed $\ln(1-p)\approx -p$, but the required small values of $p = \lambda \, dt$ can be 
reached by sufficiently small choice of the time steps $dt$. In fact, in the following considerations, we will study
the limit $dt \rightarrow 0$. Therefore, we have used the value $dt=0.001$~s in our
computer simulations.
\par
Now, let $k_i$ denote the size of the $i$th gap $T_i = t_i - t_{i-1}$
(i.e. the number of time steps $dt$ with no pedestrian arrival). Then, the expected value for the time
period until a time gap of length $\tau =N\, dt$ or greater starts is given by
\begin{eqnarray}
& & \sum_{n=0}^\infty \sum_{k_1=0}^N \dots \sum_{k_n=0}^N (k_1+1 + \dots + k_n + 1) \nonumber \\
& & \quad \times (1-p)^{k_1} p \cdot \dots \cdot (1-p)^{k_n} p \cdot (1-p)^N \nonumber \\
 &\approx & \sum_{n=0}^\infty \lambda^n 
\int\limits_0^{\rm \tau} dT_1 \dots \int\limits_0^{\rm \tau} dT_n \nonumber \\
& & \quad \times 
 (T_1 + \dots + T_n) \mbox{e}^{-\lambda (T_1+\dots +T_n)} \mbox{e}^{-\lambda \tau} \nonumber \\
&=& - \mbox{e}^{-\lambda \tau} \sum_{n=0}^\infty \lambda^n \frac{d}{d\lambda}
 \prod_{i=1}^n \bigg( \int\limits_0^{\tau} dT_i \, \mbox{e}^{-\lambda T_i} \bigg) \nonumber \\
&=&  - \mbox{e}^{-\lambda \tau} \sum_{n=0}^\infty \lambda^n \frac{d}{d\lambda}
 \left[ \frac{1}{\lambda^n} \left( 1 - \mbox{e}^{-\lambda \tau} \right)^n \right] \nonumber \\
 &=&  - \mbox{e}^{-\lambda \tau} \sum_{n=0}^\infty n \left( 1 - \mbox{e}^{-\lambda \tau} \right)^n 
\left( \frac{\tau\mbox{e}^{-\lambda \tau} }{1 - \mbox{e}^{-\lambda \tau} } - \frac{1}{\lambda} \right) \nonumber \\
 &=& \left( \frac{1}{\lambda} - \frac{\tau\mbox{e}^{-\lambda \tau} }{1 - \mbox{e}^{-\lambda \tau} }\right)
 \mbox{e}^{-\lambda \tau} s \frac{d}{ds} \sum_{n=0}^\infty s^n \mbox{ with $s=1- \mbox{e}^{-\lambda \tau}$} \nonumber \\
&=& \frac{1}{\lambda} [ \, \mbox{e}^{\lambda \tau} - (1+\lambda \tau) ]
\approx \frac{\lambda \tau^2}{2} + \dots \label{waiting}
\end{eqnarray}
That means, the average waiting time for a gap of size $\tau$ or larger starts to grow
linearly with the pedestrian arrival rate $\lambda = p/dt$ and quadratically with $\tau$ 
as long as these values are small, but it grows exponentially with $\lambda \tau$, when
this value is large.
\par
\begin{figure}
\begin{center}
\includegraphics[width=8cm]{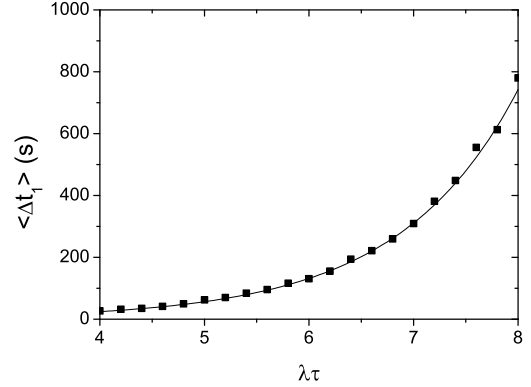}
\end{center}
\caption[]{Average waiting time $\langle \Delta t_1 \rangle$ of the first vehicle in the queue as a function
of the scaled pedestrian arrival rate $\lambda \tau$ (symbols = numerically determined values, solid line 
= analytical formula).\label{fig6}}
\end{figure}

Note, however, that the waiting time is reduced by the gap between the time $M\, dt:= v_0/a$ when the
vehicle is stopped and the time $t_n \le t_-$ at which the last pedestrian has entered the street before. 
Analogously to Sec.~\ref{earlier}, we can calculate the expected value of this time gap as
\begin{equation}
 \langle v_0/a - t_n\rangle = (1-p)  \frac{1- (1-p)^M}{p/dt} = \frac{1-\mbox{e}^{-\lambda v_0/a}}{\lambda} \, ,
\end{equation}
since we have $(1-p) \rightarrow 1$ in the limit $dt \rightarrow 0$.
As a consequence, the
expected value $\langle \Delta t_1 \rangle$ of the time period $\Delta t_1$ the first vehicle in the queue
has to wait can be estimated as
\begin{eqnarray}
 \langle \Delta t_1 \rangle &=& \frac{1}{\lambda} [ \, \mbox{e}^{\lambda \tau} - (1+\lambda \tau) ]
 - \frac{1-\mbox{e}^{-\lambda v_0/a}}{\lambda} \nonumber \\ 
  &=& \frac{1}{\lambda} \Big( \mbox{e}^{\lambda \tau} 
 +\mbox{e}^{-\lambda v_0/a} - 2 -\lambda \tau \Big)  
\end{eqnarray}
(see Fig.~\ref{fig6}).
\subsection{Average Delay to Pedestrians}

\begin{figure}
\begin{center}
\includegraphics[width=8cm]{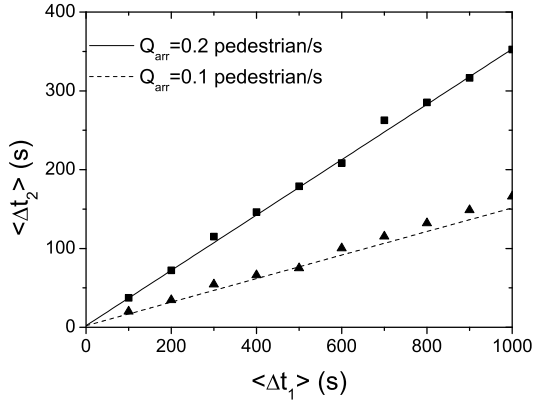}
\end{center}
\caption[]{Average time $\langle \Delta t_2 \rangle$ needed to dissolve a vehicle queue as a function
of the average time $\langle \Delta t_1 \rangle$ for which the first vehicle has been waiting, for various
values of the vehicle arrival rate $Q_{\rm arr}$, see formula (\ref{propagates}) 
(symbols = numerically determined values, straight lines 
= analytical results).\label{fig7}}
\end{figure}

After a time interval $\Delta t_1$, i.e. a time period $\tau$ after the last pedestrian has entered
the road, the first vehicle in the queue can accelerate again. The time period available to pedestrians
for crossing the road is $\Delta t_1 + v_0/a$, as the time period $v_0/a$ required to stop the vehicle
is usable as well. When the vehicle has started to move again, no pedestrian will be able to
cross the road until the last vehicle of the queue has passed point O
(at least if $\sigma \tau > T$). This time period can be calculated as \cite{control}
\begin{equation}
\Delta t_2 = C \, \Delta t_1 \frac{1+ c/v_0}{c - C} \, . 
\end{equation} 
The expected value of $\Delta t_2$ is
\begin{equation}
 \langle \Delta t_2 \rangle = C \, \langle \Delta t_1 \rangle \frac{1+ c/v_0}{c - C}  + \sqrt{ \frac{2d_0}{a}} 
\end{equation}
(see Fig.~\ref{fig7}),
where we have also taken into account the additional amount $\sqrt{2d_0/a}$ required by a vehicle 
to get from $x = -d_0$ to point O. 
\par
After the last vehicle in the queue has passed, pedestrians have a chance to find a
suitable gap of size $\sigma \tau$ or larger. A lower bound of
the expected waiting time $\langle \Delta t_3 \rangle$ for the 
occurence of such a gap is calculated in the Appendix.
In Fig.~\ref{fig8}, we compare the resulting expression
\begin{equation}
\langle T'_> \rangle = \frac{1}{Q_{\rm arr}} 
\left[ \frac{\mbox{e}^{Q_{\rm arr}(T_* -T_0)}}{1-Q_{\rm arr}T_0}
- (1+Q_{\rm arr}\, T_* ) \right] 
\label{solid}
\end{equation}
corresponding to Eq.~(\ref{erg}) with numerical results,
where $T_0=1/Q_{\rm out} = T + (l_0+d_0)/v_0$ and $T_* = \sigma \tau + (l_0+d_0)/v_0$.
This formula gives the expected waiting time $\langle \Delta
t_3 \rangle$ 
provided that the pedestrian
arrives exactly at the time, when the last vehicle in the queue
passes the crossing point. Otherwise, it is an approximation, which neglects 
\begin{enumerate}
\item the effect that pedestrians tend to arrive at the sidewalk {\em between} two vehicles
(so that there is an incomplete intervehicle time gap, which must be added), 
\item the gaps of vehicles approaching the last, already accelerating vehicle in a queue may
be smaller than $T_0$.
\end{enumerate}
These two  effects
increase the waiting time, i.e. $\langle \Delta T'_> \rangle \le \langle \Delta t_3 \rangle$. 
\par\begin{figure}
\begin{center}
\includegraphics[width=8cm]{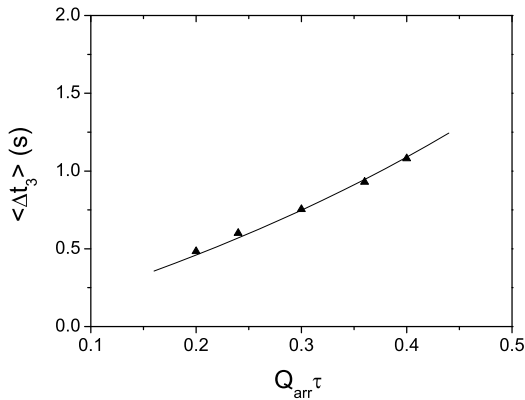}
\end{center}
\caption[]{Average waiting time $\langle \Delta t_3\rangle$ until a pedestrian enters
the road after a vehicle queue has completely dissolved, as a function of the scaled
vehicle arrival rate $Q_{\rm arr}\tau$ for $\sigma = 1.05$.
Our numerical simulation assumes the special case that
a pedestrian arrives just at time $t=t_0$ when the last
vehicle of the queue passes the crossing point.
The average time delay to this pedestrian is represented by triangles and compared
to the analytical results of formula (\ref{solid}) (solid line).\label{fig8}}
\end{figure}
During the waiting time $(\Delta t_2 + \Delta t_3)$ of pedestrians,
the expected number of arriving pedestrians is
 $\lambda \, (\Delta t_2+\Delta t_3)$. 
According to our model, all of these pedestrians will use the next occuring
gap of size $\sigma \tau$ or larger to cross the street.
We can assume that the waiting time of the last crossing pedestrian 
is approximately zero, while it is approximately $(\Delta t_2 + \Delta t_3)$ for the first one
(when the pedestrian arrival rate is high enough).
Therefore, the average delay can be approximated as $(\Delta t_2 + \Delta t_3)/2$, and the cumulative 
delay of all waiting pedestrians amounts to 
\begin{equation}
 \frac{\lambda (\Delta t_2 + \Delta t_3)^2}{2} \, .
\end{equation}

\section{Summary and Discussion} \label{summa}

In this paper, we have proposed the continous-in-space constant-deceleration-delayed-acceleration 
car-following model (CDDA model), in order to allow for the analytical calculation of 
the interactions between vehicles and crossing pedestrians under conditions of
statistically distributed arrival times. Although the model is not made to reproduce
all currently known properties of traffic flows, it does reflect some essential features
such as accident-free driving, a constant outflow from traffic jams and a characteristic
queue resolution speed. 
\par
We have distinguished two interaction modes between pedestrians and vehicles:
(i) When pedestrians prefer large safety factors $\sigma > \sigma_0$, vehicles are not
stopped, and pedestrians cross between moving vehicles either one by one or in small groups.
(ii) When pedestrians keep small safety factors $\sigma < \sigma_0$, they may stop vehicles, 
which usually causes vehicle queues. Once a large enough gap between successive pedestrian arrivals
occurs, cars will move again and prevent the crossing of pedestrians, until the last vehicle in the
queue has passed the crossing point. This oscillatory dynamics with alternating flows of
cars and pedestrians tends to be inefficient and related with long waiting times \cite{numerics}.
\par
In this contribution, we have calculated the threshold $\sigma_0$ between 
the oscillating and non-oscillating regime. It turned out to be a function of $d_0/(a\tau^2)$
only, i.e. independent of the pedestrian or vehicle arrival rates $\lambda$ and $Q_{\rm arr}$,
the vehicle length $l_0$, the free vehicle velocity $v_0$ or the preferred time gap $T$,
while the car deceleration $a$, the desired minimum distance $d_0$ and the crossing time $\tau$
matter. We have also calculated the expected waiting times of pedestrians and vehicles as a function
of the arrival rates. The difficult step in gaining these results was the calculation of 
the first overcritical time gap and its expected value. This also required the determination of the vehicle gap
distribution for deterministic, i.e. non-fluctuating vehicle interactions, while variations
in the arrival times were taken into account [see Eq.~(\ref{resu})]. The formulas for the waiting time distributions
can serve to judge under which conditions pedestrian and vehicle streams should be controlled
(terminated) by traffic lights and when a self-organized crossing of streets is more
efficient. Beyond this, our approach is generally expected to be useful for a 
better understanding of intersecting flows and certain conflicting processes.
For example, a similar gap acceptance problem is found in
lane-changing maneuvers, so that our formulas may help to
calculate analytical formulas for lane-changing rates.
\par
Regarding the choice of the behavior and parameters of cars drivers (careful or aggressive) and 
pedestrians (careful or daring), one may assume an evolutionary perspective: 
Due to a learning process during many vehicle-pedestrian interactions, an optimal
behavior should emerge on the long run. It is, however, not yet clear whether 
there exists a state which is optimal for both, drivers and pedestrians. If not, one may
consider the pedestrian-vehicle interactions as an example for a social dilemma \cite{Huberman},
and the outcome may depend on details of the interactions.
For example, if pedestrians would tend to use safety factors $\sigma < \sigma_0$, car drivers
may react to this by an aggressive approaching behavior. This would make it difficult for pedestrians
to stop vehicles. However, cars could still be successfully stopped, if pedestrians learned to enter a road exactly 
with a time gap of $\sigma_2\tau$. In conclusion, there are always strategies to
produce or avoid alternating pedestrian and vehicle flows, but the outcome depends
always on the parameters of both, pedestrian and driver behavior. The determination
of the optimal behavioral parameters and the evaluation of interactive parameter adaptations of
pedestrians and vehicles will be left for a future study.

\subsection*{Acknowledgements} 

The authors thank for partial financial support by 
the Chinese National Natural Science Foundation (Grant No. 10404025 and 10272101), 
the Alexander von Humboldt Foundation, and the German Research 
Foundation (DFG project He2789/7-1). D.H. is also grateful for inspiring discussions
with Moez Draief during the EU EXYSTENCE Thematic Institute on ``Information and Material
Flows in Complex Networks'' at Goldrain Castle, Italy.

\appendix

\section{Calculation of the expected waiting time for a suitable gap}\label{app}

Let $P(T')$ be the distribution density function of 
vehicle time gaps $T'$. Moreover, let
\begin{equation}
 Q = \mbox{Prob}(T'\le T_*) = \int\limits_0^{T_*} dT' \; P(T')
\end{equation}
be the probability of finding a time gap $T'\le T_*$ and 
\begin{equation}
 \overline{T'} := \langle T' \rangle_{T' < T_*} 
= \frac{1}{Q} \int\limits_0^{T_*} dT' \; T' P(T')
\end{equation}
the expected value of time gaps that are smaller than $T_*$. 
Then, given that a car has just passed, 
the expected time until the first gap $T'$ greater than $T_*$ occurs,
is given by the expression
\begin{equation}
 \langle T'_> \rangle = \sum_{n=0}^\infty n \overline{T'} Q^n (1-Q) \, ,
\end{equation}
as an arbitrary number $n$ of smaller gaps may occur with probability $Q$ each,
before a large enough gap occurs with probability $(1-Q)$. Here, we have used that
the expected lengths $\overline{T'}$ of short gaps $T'\le T_*$
just add up due to the assumption of independently
and identically distributed time gaps $T'$. One can calculate
\begin{eqnarray}
 \langle T'_> \rangle &=& (1-Q)\overline{T'} \bigg( \sum_{n=0}^\infty n Q^n \bigg) \nonumber \\
 &=& (1-Q)\overline{T'}  \bigg( Q \frac{d}{dQ} \sum_{n=0}^\infty Q^n \bigg) \nonumber \\
 &=&  (1-Q)\overline{T'}  Q \frac{d}{dQ} \left( \frac{1}{1-Q} \right) \nonumber \\
 &=& \frac{Q\overline{T'}}{1-Q} \nonumber \\
 &=& \frac{\int\limits_0^{T_*} dT' \; T' P(T')}
{\int\limits_{T_*}^\infty dT' \; P(T')} \, .
\label{erg1}
\end{eqnarray}
Inserting the vehicle time gap distribution (\ref{resu}) eventually gives
\begin{equation}
 1-Q = P(T'>T_*) = (1-Q_{\rm arr}T_0) \mbox{e}^{-Q_{\rm arr}(T'-T_0)} 
\end{equation}
and
\begin{equation}
 Q\overline{T'} = \frac{1}{Q_{\rm arr}} \!\left[ 1 
 - \! (1-Q_{\rm arr}T_0)(1+Q_{\rm arr}\, T_*) \mbox{e}^{-Q_{\rm arr}(T_*-T_0)}  \right] 
\end{equation}
This implies
\begin{equation}
 \langle T'_> \rangle = \frac{1}{Q_{\rm arr}} 
\left[ \frac{\mbox{e}^{Q_{\rm arr}(T_*-T_0)}}{1-Q_{\rm arr}T_0}
- (1+Q_{\rm arr}\, T_* )\right] \, .
\label{erg}
\end{equation}
The required minimum time gap for the crossing of a pedestrian 
between two successive vehicles is $T_* = \sigma \tau + (l_0+d_0)/v$,
while the preferred time gap between successive vehicles is $T_0 = T + (l_0+d_0)/v$. 
Note that formula (\ref{waiting}) for the expected waiting time of vehicles
for a large enough gap in the pedestrian stream
corresponds to the special case $T_0 =0$ with
$\lambda =Q_{\rm arr}$ and $T_* = \tau$.
\clearpage
\end{document}